\documentclass[nofootinbib, reprint, superscriptaddress, 
]{revtex4-2}

\usepackage{graphicx}
\usepackage{dcolumn}
\usepackage{amsmath}
\usepackage{braket}
\usepackage{bm}
\usepackage{color}
\usepackage{siunitx}
\usepackage{soul}
\usepackage{amssymb}
\usepackage{ulem}[normal]
\usepackage{cancel}

\usepackage{amssymb}

\newcommand{\be}{\begin{equation}}
\newcommand{\ee}{\end{equation}}
\newcommand{\beq}{\begin{eqnarray}}
\newcommand{\eeq}{\end{eqnarray}}
\newcommand{\ba}{\[\begin{aligned}}
\newcommand{\ea}{\end{aligned}\]}
\newcommand{\bal}{\begin{aligned}}
\newcommand{\eal}{\end{aligned}}

\renewcommand{\epsilon}{\varepsilon}

\renewcommand{\figurename}{Fig.}
\renewcommand{\thefigure}{\arabic{figure}}

\def \ba{\begin{align*}}
\def \ea{\end{align*}}

\newcommand{\cavnbn}{KIBC} 
\usepackage{xcolor}
\newcommand{\apnbn}{KIBC} 
\newcommand{\apg}{SNGP}

\newcommand{\SIref}{Supplementary Information}

  {}

\begin{document}

\title{    Strongly coupled ultracompressed self-cavity plasmon polaritons at millikelvin temperatures}
\begin{abstract}
Van der Waals heterostructures exhibit self-cavity resonances from sub-terahertz to terahertz frequencies and thus offer an attractive route toward realizing cavity-modified phases of matter.
Spectroscopy of such systems calls for high frequency resolution at millikelvin temperatures, which is extremely challenging with existing methods.
Here we demonstrate a new platform for realizing and probing such systems.
Superconducting layers define self-cavity structures with electromagnetic volume compression of order $10^7$. 
With an integrated graphene layer, we observe underdamped cavity plasmon polaritons at frequencies as low as 0.04 THz and which exhibit avoided crossings with cavity modes. 
To measure such spectra, we deploy a millikelvin-compatible sub-THz spectrometer based on a superconducting circuit. 
This work paves the way for simultaneous spectroscopy and transport measurements of cavity-matter hybrid systems at millikelvin temperatures.
\end{abstract}

\author{Luojia Zhang$^{*}$$^{\dagger}$}
\affiliation{Department of Physics, Cornell University, Ithaca, NY, USA}
\author{Andrew T. Pierce$^{*}$$^{\dagger}$}
\affiliation{Department of Physics, Cornell University, Ithaca, NY, USA}
\affiliation{Kavli Institute for Nanoscience at Cornell, Ithaca, NY, USA}
\affiliation{Max Planck Institute for Structure and Dynamics, Hamburg, Germany}
\author{Xuepeng Wang}
\affiliation{Department of Physics, Cornell University, Ithaca, NY, USA}
\author{Simon Reinhardt}
\affiliation{Department of Applied and Engineering Physics, Cornell University, Ithaca, NY, USA}
\author{Kenji Watanabe}
\affiliation{Research Center for Electronic and Optical Materials, National Institute for Materials Science, Tsukuba, Japan}
\author{Takashi Taniguchi}
\affiliation{Research Center for Materials Nanoarchitectonics, National Institute for Materials Science, Tsukuba, Japan}
\author{Kin Fai Mak}
\affiliation{Department of Physics, Cornell University, Ithaca, NY, USA}
\affiliation{Max Planck Institute for Structure and Dynamics, Hamburg, Germany}
\affiliation{Department of Applied and Engineering Physics, Cornell University, Ithaca, NY, USA}
\author{Jie Shan}
\affiliation{Department of Physics, Cornell University, Ithaca, NY, USA}
\affiliation{Max Planck Institute for Structure and Dynamics, Hamburg, Germany}
\affiliation{Department of Applied and Engineering Physics, Cornell University, Ithaca, NY, USA}
\author{Debanjan Chowdhury}
\affiliation{Department of Physics, Cornell University, Ithaca, NY, USA}
\author{Valla Fatemi$^\dagger$}
\affiliation{Department of Applied and Engineering Physics, Cornell University, Ithaca, NY, USA}

\date{\today}

\maketitle

\def\thefootnote{*}\footnotetext{These authors contributed equally to this work.}
\def\thefootnote{$\dagger$}\footnotetext{Corresponding authors' emails: lz282@cornell.edu (L.Z.), atp66@cornell.edu (A.T.P.), vf82@cornell.edu (V.F.).}
\def\thefootnote{\arabic{footnote}}
\makeatletter
\renewcommand{\figurename}{\textbf{Fig.}}
\def\fnum@figure{\textbf{\figurename~\thefigure}}
\makeatother



Electromagnetic cavities with integrated quantum materials offer opportunities for realizing hybridized light-matter systems and cavity-modified ground states \cite{ciuti_quantum_2005, khitrova_vacuum_2006, sentef_cavity_2018, fatemi_synthesizing_2018,  flick_strong_2018, li_vacuum_2018, garcia-vidal_manipulating_2021,thomas_large_2021, schlawin_cavity_2022, appugliese_breakdown_2022,jarc_cavity-mediated_2023,passetti_cavity_2023,lu_cavity-enhanced_2024, hubener_quantum_2024,eckhardt_surface-mediated_2025, helmrich_cavity-driven_2025, enkner_tunable_2025, xu_vacuum-dressed_2026, keren_cavity-altered_2026, bretscher_fluctuation_2026}.
Van der Waals heterostructures are an attractive platform for such systems due to their high quality and rich materials landscape \cite{basov_polaritons_2016, low_polaritons_2017, gallagher_quantum-critical_2019,  kennes_moire_2021, mak_semiconductor_2022, seo_-chip_2024, michael_resolving_2025, kipp_cavity_2025, michael_resolving_2025, chen_terahertz_2025}.
However, conventional cavity-matter architectures face several challenges. 
First, the small length scale of van der Waals heterostructures necessitates near-field probes operating in the tens to hundreds of GHz, beyond the range of conventional microwave spectroscopy.
Second, plasmonic cavities based on non-superconducting materials are intrinsically lossy, greatly limiting their quality factors and thus their sensitivity to the collective modes of the materials.
Third, operation at millikelvin temperatures is essential for stabilizing the most delicate quantum states, 
complicating the use of spectroscopic probes based on optics.

Here we present a new platform combining two key components for realizing and probing highly confined van der Waals cavities in the sub-THz frequency range  (Fig. 1a).
The first component is a  novel cavity geometry consisting of two layers of a high kinetic inductance (KI) superconductor (SC) arranged in a parallel-plate or ``broadside'' geometry. 
This structure, which we dub the kinetic inductance broadside cavity (KIBC), supports ultra-slow plasmonic propagation velocities, of order $100 \times$ below that of free space. 
Incorporating a quantum material inside the cavity then adds an additional plasmon mode that directly reflects the dynamic conductivity of the quantum material.
Together, the two plasmonic modes in the combined structure respectively serve as ``light'' and ``matter'' degrees of freedom (Fig. 1b).

The second key component of our platform is a frequency-domain, on-chip, sub-THz spectrometer based on Josephson junctions \cite{hofheinz_bright_2011,bretheau_exciting_2013, griesmar_superconducting_2021, peugeot_two-tone_2024} 
that allows us to perform spectroscopy in the weak-perturbation regime up to nearly 200 GHz and in the quantum limit ($k_B T \ll h f$).
Compared to optics-based, time-domain methods for generating and receiving sub-THz signals, the Josephson spectrometer reduces heat loads and peak electric fields by several orders of magnitude  (Fig. 1c), while also improving frequency resolution by up to two orders of magnitude.
Taken together, these components offer the first viable approach to probing vdW-based cavity quantum matter at THz frequencies and at mK temperatures.


\begin{figure*}
\begin{center}
\includegraphics[width=0.95\textwidth]{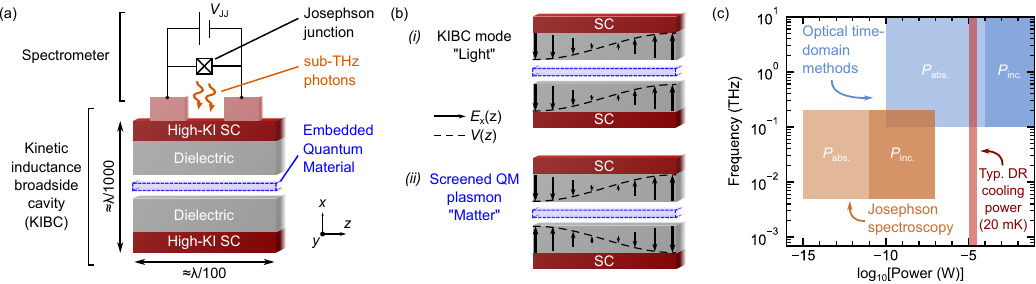}
\end{center}
\caption{\textbf{Kinetic inductance broadside cavities (KIBCs) probed by Josephson spectroscopy.} 
\textbf{(a)} Schematic side view of a KIBC coupled to a Josephson spectrometer. 
The KIBC is defined by high-KI superconductors separated by dielectric layers. 
Low-dimensional quantum materials can be sandwiched between the two layers. 
The cavity is connected to a Josephson junction that acts as an emitter and detector of sub-THz photons, allowing for measurement of the high-frequency electrodynamic response of the KIBC.
\textbf{(b)} Structure of the plasmonic modes of a trilayer structure. 
Solid arrows label the orientation of the transverse electric field $E_x$.
Dashed lines indicate the profile of the interlayer voltages.
In the KIBC ``light'' mode, the electric fields above and below the middle layer point in the same direction; for the screened plasmon of the quantum material (``matter'') mode they point in opposite directions.
\textbf{(c)} Comparison between the available frequency range and dissipated power for Josephson spectroscopy and optics-based time-domain THz methods. 
The cooling power at 20 mK for typical dilution refrigerators is highlighted in red \cite{bluefors_note}.
$P_{\text{inc.}}$ is defined for the time-domain scheme as the typical required incident optical power on the sample, and for JJ spectroscopy as the power dissipated at the mixing chamber.
$P_{\text{abs.}}$ is defined as the typical absorbed sub-THz power.
The parameters used in the comparison are given in the \SIref.
}
\label{fig:intro}
\end{figure*}


In this manuscript, we describe our results on a sequence of three cavity-matter structures of progressively increasing complexity.
First, we discuss a superconductor-superconductor bilayer cavity i.e., a bare KIBC, with which we establish high-quality KIBC modes. 
Second, we consider a superconductor-graphene bilayer (hosting screened graphene plasmons, SNGPs), and observe notably long plasmonic lifetimes that appear to be limited by the size of the device.
Finally, we describe a superconductor-graphene-superconductor trilayer that hybridizes KIBC modes and SNGP modes, finding strong coupling in the form of avoided crossings. 
These observations are made possible by the Josephson spectrometer, which we here apply to vdW systems for the first time: the sub-GHz frequency resolution enables us to observe couplings down to the 1-2 µeV level, far below the resolution of optics-based methods.
Overall, our findings establish a new approach to realizing and probing cavity-matter structures at ultralow temperatures, opening the door to future studies of hybrid light-matter systems.

\section*{Josephson Dynamical Conductivity Spectroscopy for Condensed Matter Systems}

We first describe the Josephson spectrometer and its application to vdW systems at mK temperatures.
Josephson spectroscopy uses a Josephson tunnel junction (JJ) as a simultaneous emitter and detector of sub-THz radiation \cite{hofheinz_bright_2011,bretheau_exciting_2013, griesmar_superconducting_2021,peugeot_two-tone_2024}.
The emission comes from the AC Josephson effect, in which a nonzero DC voltage bias $V_{\text{JJ}}$ across the JJ results in an oscillating AC current $I(t) \approx I_0 \sin (\omega_{\text{J}} t)$ across the JJ at the frequency $\omega_{\text{J}}/2 \pi=2eV_{\text{JJ}}/h \approx 483.6 \times V_{\text{JJ}}$ GHz/mV, where $I_0$ is the critical current of the JJ. 
The maximum emission frequency is $4 \Delta / h$, where $\Delta$ is the quasiparticle gap of the superconductor, beyond which the junction transitions to the resistive state. 
For standard thin film Al JJs, this is roughly 190 GHz. 

The detection process comes from inelastic Cooper pair tunneling across the same JJ: if the local electromagnetic environment dissipates some of the emitted radiation at $\omega_{\text{J}}$, then Cooper pairs tunnel inelastically across the junction at DC to supply the necessary power \cite{devoret_effect_1990, ingold_finite-temperature_1991, holst_effect_1994, deblock_detection_2003, hofheinz_bright_2011,  grimm_bright_2019}. 
Quantitatively,  the DC current is determined by the real part of the input impedance $Z_{\text{in}}(\omega_{\text{J}})$ via $I_{\text{DC}}=I_0^2 \text{Re} \left[ Z_{\text{in}}(\omega_{\text{J}})\right]/2V_{\text{JJ}}$.
Because the emission spectrum is narrow-band, $I_{\text{DC}}$ serves as a frequency-sensitive probe of the dissipation of the JJ's environment.
Thus, the Josephson spectrometer has a key advantage in only requiring DC biasing, which substantially reduces complexity and cost relative to direct use of sub-THz technology.


\begin{figure*}
\begin{center}
\includegraphics[width=0.95\textwidth]{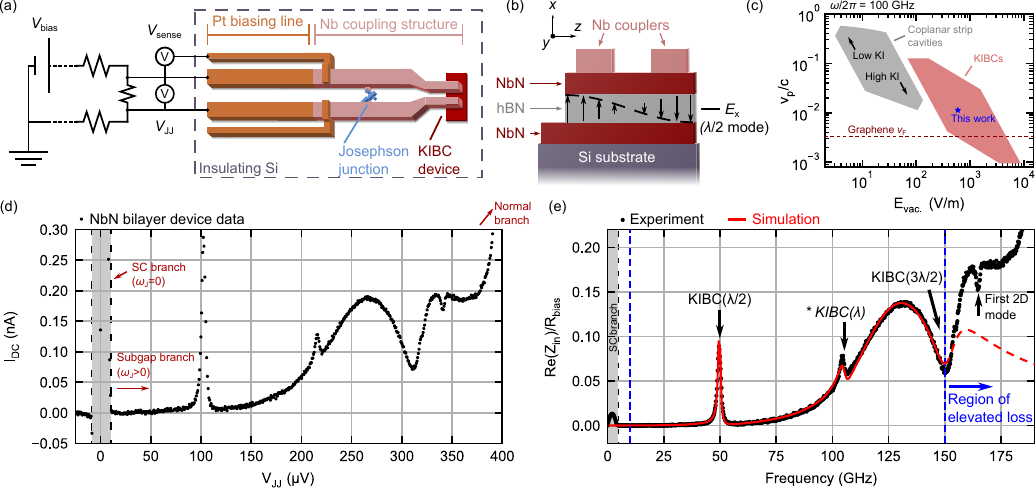}
\end{center}
\caption{\textbf{Josephson spectroscopy of a superconductor/superconductor \apnbn. } 
\textbf{(a)} Simplified schematic of the Josephson spectroscopy circuit. 
An insulating Si substrate hosts an aluminum Josephson junction connected to Nb electrodes, which are in turn connected to a lossy Pt transmission line. 
The Pt transmission line is wire bonded to an on-chip voltage divider on a second, separate Si substrate (see \SIref). 
The voltages $V_{\text{sense}}$ and $V_{\text{JJ}}$ are probed by measuring across the Pt transmission line electrodes, and the DC junction current is given by $V_{\text{sense}}/R_{\text{bias}}$. 
\textbf{(b)} Side view of the NbN/hBN bilayer \apnbn. 
\textbf{(c)} Normalized velocity $v_\text{p}/c$ versus vacuum electric field $E_\text{vac.}$ for KIBCs compared to coplanar strip-type cavities, including the devices used in this work. 
We consider the out-of-plane field component for the KIBC modes and the longitudinal component for the CPS modes; the longitudinal component of the KIBC modes is similar to that of the CPS modes.
The parameters used in the comparison are given in the \SIref.
\textbf{(d)} Measured $I$-$V$ characteristic of a Josephson spectrometer coupled to an NbN KIBC. 
Spectroscopy is performed by fixing the junction in the subgap (finite-voltage) state. 
Peaks in the subgap branch correspond to the \apnbn~modes. 
\textbf{(e)} Extracted $\text{Re}\left[Z_{\text{in}}( \omega )\right]$ obtained from the data in panel d (see \SIref). 
The \apnbn~modes appear peaks and dips in the spectrum. 
The blue dashed lines indicate the region over which the global fit is performed. 
The mode frequencies correspond to a plasmon velocity of approximately $v_{\text{p}} = 3.4 \times 10^6$ m/s.
}
\label{fig:spectrometer_and_APNbN}
\end{figure*}

A schematic illustration of the spectrometer device is shown in Fig. 2a. 
The JJ is connected to Nb electrodes and is biased through a resistive Pt transmission line. 
These structures are engineered to give specific RF impedance characteristics, over several orders of magnitude in frequency.
The samples we probed were placed at the end of the Nb electrodes and coupled through overlapping Nb couplers. 
Finally, we remark on the active and passive power requirements. 
On-chip AC powers are 
at the fW to pW scale, most of which is dissipated in the Pt biasing lines.
Overall, these power loads are manageable at millikelvin temperatures \cite{blagg_-chip_2022, bon-mardion_sub-kelvin_2026},
and may be compared with complementary optics- and time-domain-based THz spectroscopy systems 
and with typical dilution refrigerator cooling powers (Fig.~\ref{fig:intro}c).
We note that the effective cooling power delivered to the electrons in a van der Waals device is difficult to estimate, but is expected to be orders of magnitude lower than the total dilution refrigerator cooling power.
All measurements are performed at the base temperature $T_{\text{base}} \approx 10$ mK of a dilution refrigerator with no detectable rise in mixing chamber temperature during operation of the spectrometer.
The spectrometer is described in more technical detail in the \SIref.

\section*{Ultracompact superconducting plasmonic cavities}

We first describe the results of the experiment on the simplest KIBC-type structure formed from two NbN plates separated by a thin hBN dielectric layer (Fig. 2b). 
We used disordered superconducting NbN films made by magnetron sputtering for the conductor layers,
which
exhibit high kinetic inductance $L_{\text{k}}$ (corresponding to low superfluid stiffness $\rho_{\text{s}} \propto L_{\text{k}}^{-1}$) \cite{potts_-chip_2023, potts_finite-momentum_2026}, while maintaining high critical temperature $T_c$. 

Combining these components yields plasmonic KIBC modes with linear dispersion relations, sometimes referred to as ``acoustic'' plasmon modes \cite{hwang09, alonso-gonzalez_acoustic_2017, barcons_ruiz_experimental_2023}. 
In our system, these modes are characterized by low phase velocities $v_\text{p} \propto (\epsilon L_{\text{k}}/d)^{-1} \approx c/100$ (where $d$ and $\epsilon$ are the thickness and permittivity of the dielectric), high quality factors, and moderate characteristic impedances.
With these velocities, cavities of a few tens of microns host cavity modes at sub-THz frequencies. 
Alongside the slow plasmon velocities, the high KI layers 
enable out-of-plane vacuum electric field fluctuations about two orders of magnitude larger than longitudinal fluctuations in typical coplanar resonators (Fig. 2c). Thus, KIBC modes are especially suitable for coupling to displacement field-tuned phases and excitations.

The sub-gap portion of the current-voltage characteristic of a Josephson spectrometer coupled to a typical NbN \cavnbn~is shown in Fig. 2d. 
The data exhibit several resonances, which we identify as the \apnbn~modes of the NbN bilayer. 
To make a quantitative comparison with theory, we extracted the normalized input impedance $\text{Re}(Z_{\text{in}})/R_{\text{bias}}$ from the $I$-$V$ characteristic (see \SIref), yielding the spectrum shown in Fig. 2e. 
The most prominent features in the spectrum are a sharp peak at 50 GHz and a large dip around 150 GHz. 
We identified these features as the $\lambda/2$ and $3\lambda/2$ \apnbn~modes in agreement with the expected mode frequencies for the geometry of the device.  
A smaller peak near 100 GHz corresponds to the $\lambda$ mode, which is less visible due to the nearly-symmetric placement of the couplers on the sample.
Finally, the small dip near 160 GHz is consistent with a mode whose wavevector has components along both $z$ and $y$ (see \SIref). 

We simulated the response of the \cavnbn~using an effective circuit model.
We allowed for a shunt resistance-type loss on the NbN layers. 
Next, we used a model of the full spectrometer circuit
to calculate the total impedance.
We performed a multi-parameter fit to the region between the dashed lines in Fig. 2e and obtain a simulated $Z_{\text{in}}$ that agrees quantitatively with the experimental data (red curve, Fig. 2e). 
Full details 
are given in the \SIref. 
Our model successfully accounts for all 
modes observed up to 150 GHz.

Overall, these findings demonstrate low-loss superconducting sub-THz cavities based on \cavnbn~modes, setting the stage for coupling
to van der Waals materials. 
In addition to their strong confinement and vacuum fields, the \cavnbn~modes in Fig. 2e indicate remarkably low phase velocities corresponding to approximately $3.4 \times 10^6$ m/s.
Thus,
ultracompact plasmon modes with phase velocities approaching the Fermi velocities of metallic materials, including graphene ($v_{\text{F}} \approx 10^6$ m/s), can be readily achieved. 

\section*{Ultraslow nonlocal graphene plasmons}
\begin{figure}
\begin{center}
\includegraphics[width=0.45\textwidth]{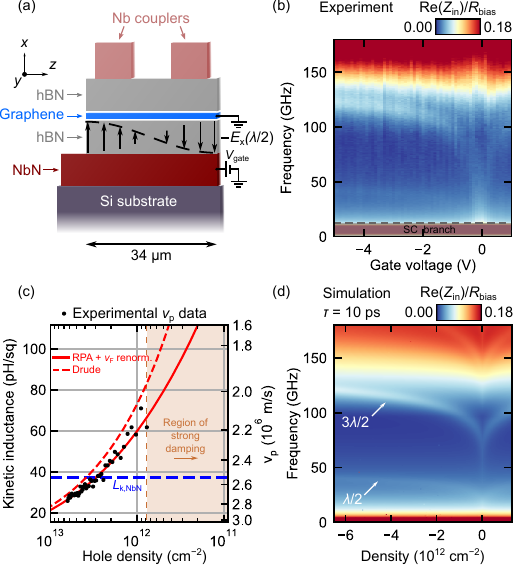}
\end{center}
\caption{\textbf{Long-lived hybrid graphene/high-KI superconductor plasmons.} 
\textbf{(a)} Schematic of NbN-graphene bilayer plasmonic cavity structure. 
\textbf{(b)} Experimental Josephson impedance spectra of an NbN/graphene bilayer device measured as a function of density.
\textbf{(c)} RPA-corrected KI of graphene as a function of density compared with that of NbN. 
The corresponding mode velocity is shown on the right axis. 
The red curves are obtained from simulations using the effective circuit model. 
The black markers correspond to velocities extracted from the data in panel b (see text).
$L_{k,\text{NbN}}=37$ pH/sq and $\epsilon_{\text{hBN}}=3.5 \epsilon_0$ are found to give good agreement between the RPA simulation and the experimental data points.
\textbf{(d)} Simulated input impedance of an NbN/graphene bilayer using the circuit model discussed in the text. 
The two modes visible in the experiment are identified as the $\lambda/2$ and $3 \lambda/2$ cavity modes. The relaxation time is set to 10 ps.
}
\end{figure}
\label{fig:AP_GNbN}
To demonstrate the possibility of probing confined collective modes of quantum matter, we turn to a bilayer device consisting of one NbN layer and one monolayer graphene layer (Fig. 3a) \cite{lundeberg_tuning_2017, torre_acoustic_2019, barcons_ruiz_experimental_2023},
which hosts a density-tunable hybrid graphene/SC mode we refer to as 
a screened nonlocal graphene plasmon (SNGP) mode.

We performed Josephson spectroscopy on the device shown schematically in Fig. 3a, resulting in the normalized impedance spectrum depicted in Fig. 3b.
Our device includes contacts to the NbN and graphene layers that allow the graphene carrier density to be tuned by the electrostatic field effect.
A prominent mode near 120 GHz at high densities ($V_\text{gate} \approx -5$ V) 
drops sharply 
as the gate voltage is reduced, and eventually broadens considerably at charge neutrality. 
A weaker mode 
near 40 GHz 
also redshifts and gradually disappears as the density is reduced.
We respectively identified the 40 and 120 GHz modes as the $\lambda/2$ and $3\lambda/2$ SNGP modes.

To understand the variation of the mode frequencies with density, we note that the KI of monolayer graphene can be tuned by varying the carrier density, and the long scattering times possible for graphene suggest that this reactive response can be dominant~\cite{lundeberg_tuning_2017, ni_fundamental_2018, kipp_cavity_2025}. 
We modeled the mode velocity by adapting the effective-circuit formalism to incorporate the nonlocal
$q$-dependent
electrodynamic response of graphene \cite{wunsch_dynamical_2006, hwang_dielectric_2007, polini_plasmons_2008, lundeberg_tuning_2017}  along with the expected Fermi velocity renormalization at low density due to interaction effects \cite{barlas_chirality_2007, abedinpour_drude_2011, das_sarma_velocity_2013, yu_interaction_2013, lundeberg_tuning_2017}.
These effects were included via the identification $\sigma_\text{sheet} \rightarrow i \omega e^2/q^2 \times \Pi(q, \omega) $, where $\Pi(q,\omega)$ is the polarizability of a single graphene layer calculated within the RPA 
\cite{pines_1, pines_2, pines_3, Mahan1990ManyParticle, pines1989theory, Giuliani:826125, sarma_07, rmp_graphene}
and relaxation-time approximations \cite{mermin_lindhard_1970, lundeberg_tuning_2017} and with the renormalization of $v_{\mathrm{F}}$ taken into account. 
This identification is justified by a remarkable correspondence between the RPA saddle-point equation of motion and the classical wave equation of a coupled multilayer transmission line (see \SIref). 

Fig. 3c depicts the effective KI of monolayer graphene as a function of carrier density including self-consistent-field corrections and $v_{\mathrm{F}}$ renormalization. 
For densities below approximately $2 \times 10^{12}$ cm$^{-2}$, the KI of monolayer graphene exceeds that of the NbN film, resulting in a widely tunable plasmon velocity (right axis, Fig. 3c).
As shown in Fig. 3c, the plasmon phase velocity approaches (within a factor of 2) the Fermi velocity of graphene $v_\text{F} \approx 10^6$ m/s as the hole concentration is lowered.
As a result, the RPA predicts corrections on the order of 10-20\% for densities around $10^{12}$ cm$^{-2}$ and below compared to the naive $q$-independent Drude conductivity of graphene. 
To compare with the experiment, we extracted approximate plasmon velocities for $p$ doping by 
tracking the peak of the SNGP($3\lambda/2$) mode in Fig. 3c.
The velocities $\omega/q$ extracted for the corresponding $q=\pi/(34\text{ µ}\text{m})$ are plotted in Fig. 3c; we found good agreement between the RPA model and the velocity data points for $\epsilon=3.5 \epsilon_0$ and $L_{k,\text{NbN}}=37$ pH/sq. 
For densities below $8 \times 10^{11}$ cm$^{-2}$, the peak in the experimental data is too broad to be reliably identified.

We used our effective-circuit model to simulate the total input impedance $Z_{\mathrm{in}} (\omega)$ as a function of the graphene scattering time $\tau$.
We found reasonable agreement between experiment and simulations when the relaxation time $\tau$ is taken to be 10 ps (Fig. 3d); for shorter values of $\tau$ the model predicts modes that are broader in frequency than seen in the experiment
(see 
\SIref). 
We note that this plasmon lifetime is comparable to or exceeds the longest reported lifetimes in the literature \cite{ni_fundamental_2018, kipp_cavity_2025}.
In particular, we find $v_p \tau/L \approx 1$ at high doping, where $L$ is the cavity length, suggesting that the mode quality factor and lifetime are limited by edge scattering.
We note that while tunable screened graphene plasmons in the THz frequency range have been measured in the past \cite{barcons_ruiz_experimental_2023, lundeberg_tuning_2017}, 
prior realizations used highly conductive and dissipative normal metal counter-plates,
which resulted in
plasmon lifetimes lower by a factor of 30 to 40
relative to the inverse linewidths we observe.
This underscores the advantage of working at sub-THz frequencies where $\hbar \omega < 2 \Delta$.
Overall, these findings establish hybrid superconductor-graphene confined plasmonic modes as a viable platform for constructing high-quality tunable sub-THz cavities.

\section*{Strong light-matter coupling in hybrid van der Waals/high-KI SC plasmonic cavities}
Finally, we turn to the question of realizing strong coupling between the graphene plasmon and the cavity modes.
To do so, we constructed a trilayer NbN/graphene/NbN structure, as shown schematically in Fig. 4a.
This three-layer structure possesses two confined plasmon modes which are sketched in Fig. 4a(i)-(ii).
\begin{figure*}
\begin{center}
\includegraphics[width=0.95\textwidth]{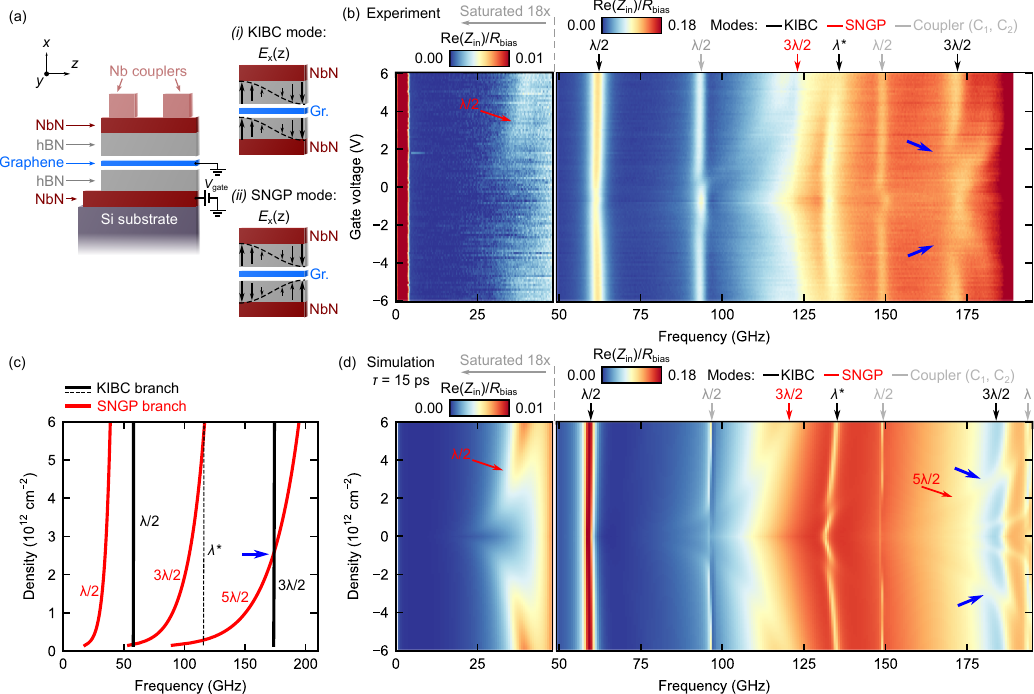}
\end{center}
\caption{\textbf{Hybrid graphene-superconducting confined plasmons and strong cavity coupling.} 
\textbf{(a)} Schematic of NbN-graphene-NbN plasmonic trilayer structure. 
\textbf{(b)} Experimental Josephson impedance spectra of an NbN/graphene/NbN trilayer device measured as a function of density. 
Frequencies below 50 GHz are saturated to highlight the weak \apg$(\lambda/2)$ mode. 
The blue arrows indicate avoided crossings (corresponding to the level crossing highlighted in Fig. 4c). 
\textbf{(c)} Dispersion of the {\apnbn}$(\lambda/2)$ and {\apg}$(\lambda/2)$ modes, along with several harmonics. 
The mode frequencies are calculated for an infinite NbN/graphene/NbN trilayer system with no Nb couplers.
$\lambda/2$ is set to 34 µm. 
The starred \apnbn$(\lambda)$ mode is formally forbidden and only appears in the presence of mirror symmetry breaking. The blue arrow indicates a level crossing. 
\textbf{(d)} Simulated input impedance of an NbN/graphene/NbN trilayer using the circuit model discussed in the text. 
Modes are highlighted in black, red and grey, and the blue arrows highlight the avoided crossings observed in the experiment. 
The relaxation time is set to 15 ps.}
\end{figure*}
\label{fig:strong_coupling}

For one mode (Fig. 4a(i)), the electric field between the top NbN layer and graphene points in the same direction as that between graphene and the bottom NbN layer. 
This electric field configuration produces minimal current flow in the graphene layer, and 
thus
the mode exhibits little dependence on the graphene KI. 
We therefore identify this mode as a \apnbn, or ``cavity'' mode.
The second mode (Fig. 4a(ii)) hosts electric fields sourced from the graphene layer, so it will be sensitive to the electrodynamic properties of graphene.
We identify this mode as the SNGP or ``matter'' mode. 
We emphasize that a mode spectrum including both ``cavity'' (KIBC) and ``matter'' (SNGP) plasmonic modes within this frequency band is straightforwardly realized with high-KI superconducting layers, whereas lossy normal-metal layers like Au would not support their own cavity modes at these frequencies.

We performed Josephson spectroscopy on the device shown schematically in Fig. 4a, yielding the normalized impedance spectrum under varying gate voltage shown in Fig. 4b.
The spectra exhibit numerous modes, some of which disperse strongly as a function of density and some of which disperse minimally.
In particular, we found minimally dispersing modes around 62, 130, and 175 GHz, with the mode at 62 GHz being the strongest sharp feature in the data.
Dispersing modes are evident near 40 GHz (left panel, Fig. 4b) and 120 GHz at high carrier density. 
A third dispersing feature is observed to cut through the nondispersing mode around 170 GHz, yielding an avoided crossing (blue arrows, Fig. 4b). 

To identify the features in the experiment, we examined the theoretical dispersion relations of the KIBC and SNGP modes.
Fig. 4c shows predictions of the first few harmonics of the {\apnbn} and {\apg} modes at $q=\pi/34$ µm, consistent with experimental cavity length of 34 µm but for an infinite, translationally symmetric system.
Comparison between Figs. 4b and 4c allowed us to identify each harmonic individually, with strongly dispersing features corresponding to SNGP modes and weakly dispersing features corresponding to KIBC modes
(labels, Fig. 4b).
Notably, it follows from the mode dispersions that a level crossing between the \apnbn$(3 \lambda/2)$ and \apg$(5 \lambda/2)$ modes is expected to occur around 170 GHz (blue arrow, Fig. 4b).
In our cavity structure, hybridization occurs at level crossings because of 
the couplers, which introduce spatial inhomogeneity.
Thus, we interpret the avoided crossing in Fig. 4b as a result of hybridization between the \apnbn$(3 \lambda/2)$ and \apg$(5 \lambda/2)$ modes, thereby demonstrating strong light-matter coupling between confined plasmonic modes in our hybrid KIBC/quantum material architecture.

The spectrum includes additional 
modes at 94 GHz and 147 GHz.
These peaks cannot be explained as harmonics of the {\apnbn} modes. 
We instead interpret these features as ``coupler modes,'' respectively labeled $C_1$ and $C_2$, arising from a thin capacitive interface between the Nb couplers and the NbN.
Intriguingly, the coupler modes also exhibit signatures of hybridization with the graphene modes with sub-GHz coupling (see \SIref), underscoring how this spectroscopic method has the resolution to reveal subtle features at µeV energy scales. 
We discuss these coupler modes in more depth in the \SIref. 

We modeled the measured impedance using our effective-circuit approach and obtained the spectrum shown in Fig. 4d.
The exact parameters used in the model are specified in the \SIref. 
Overall, the model calculation accounts for all of the modes observed in the experiment, with some discrepancies in the frequency or amplitude of some of the modes and avoided crossings. 
We attribute these discrepancies to our uncertainty in the properties of the coupler regions
We note that recently predicted cavity-induced effects on the electronic structure itself, e.g. effective mass renormalization \cite{eckhardt_surface-mediated_2025}, would also manifest in our experiments as shifts in the mode frequencies and may not be captured by our model.
KIBCs probed with Josephson spectroscopy furnish a promising approach to distinguish such ``back-action'' type effects in future experiments.

As in the case of the NbN-graphene bilayer, 
our trilayer device exhibits a surprisingly long graphene plasmon lifetime of roughly $\tau \approx 15$ ps 
(
see \SIref
), 
again consistent with $v_p^{\text{SNGP}} \tau/L \approx 1$, where $L$ is the length of the cavity. 
This reproduces the observation of SNGP mode lifetimes likely limited by the device size via edge scattering.

\section*{Conclusion}

Overall, our findings establish superconducting {\apnbn}s as a unique and effective platform for realizing strong light-matter coupling \cite{ciuti_quantum_2005, sentef_cavity_2018, flick_strong_2018, li_vacuum_2018, appugliese_breakdown_2022, hubener_quantum_2024, garcia-vidal_manipulating_2021, fatemi_synthesizing_2018, khitrova_vacuum_2006} and for probing the dynamic conductivity of low-dimensional quantum materials \cite{rmp_graphene, michael_resolving_2025, gallagher_quantum-critical_2019, seo_-chip_2024, kipp_cavity_2025, michael_resolving_2025,  chen_terahertz_2025}.
The ultra-low mode volume and low loss intrinsic to the NbN {\apnbn}s, as well as our mK-compatible spectroscopy scheme based on Josephson junctions, is especially suitable for experiments on a wide variety of low-dimensional correlated electron systems \cite{rmp_graphene, kennes_moire_2021, mak_semiconductor_2022} and raises the possibility of coupling to collective modes of such systems \cite{rmp_graphene, sun_universal_2018, sun_collective_2020, chen_terahertz_2025}.
In particular, we note that a number of collective modes tend towards a vanishing coupling matrix element in the zero momentum limit, so the high effective momentum of our KIBC structure can push into regimes where such collective modes become visible~\cite{sun_collective_2020}. 
Crucially, our spectrometer allows for simultaneous, \textit{in situ} measurements of transport properties in a single device, thus providing a means of correlating high-frequency response with ground-state phase diagrams, which we leave to future work.
Finally, we anticipate that further optimization of the dielectric thickness, dielectric constant, and kinetic inductance may enable an additional order-of-magnitude enhancement of the vacuum field amplitude, with corresponding reductions in plasmon velocity.
These advances pave the way for investigating the interplay of material phases with structured and strongly-coupled electromagnetic environments~\cite{fatemi_synthesizing_2018, schlawin_cavity_2022,  keren_cavity-altered_2026,  jarc_cavity-mediated_2023, thomas_large_2021, lu_cavity-enhanced_2024,  xu_vacuum-dressed_2026}.

\section*{Author Contributions}

L.Z. and A.T.P. designed and fabricated the spectrometer devices.
L.Z. fabricated the van der Waals devices and performed the electrical measurements. 
A.T.P. performed the circuit analysis. 
S.R. contributed the Josephson junction fabrication recipe. 
X.W. performed the RPA analysis with input from D.C. 
T.T. and K.W. provided the hBN crystals. 
V.F. supervised the project. L.Z., A.T.P. and V.F. wrote the manuscript with input from all authors.

\begin{acknowledgements}
We acknowledge assistance on developing Josephson junctions recipes from Britton L. Plourde and Gabriele Di Gianluca.
L.Z. acknowledges support from a fellowship from David W. Meehl in memory of James R. Meehl and supported by Cornell Duffield Engineering.
We gratefully thank the Aref and Manon Lahham Faculty Fellowship for support of L.Z. and this project. 
A.T.P. acknowledges support from the Kavli Institute at Cornell for Nanoscale Science (KIC) Postdoctoral Fellowship.
D.C. is supported in part by a grant from the Department of Energy (DE-SC0026112) under the Early Career Research Program.
K.W. and T.T. acknowledge support from the CREST (JPMJCR24A5), JST and World Premier International Research Center Initiative (WPI), MEXT, Japan.
This work was performed in part at the Cornell NanoScale Facility, a member of the National Nanotechnology Coordinated Infrastructure (NNCI), which is supported by the National Science Foundation (NSF) (Grant No. NNCI-2025233).
\end{acknowledgements}

\bibliography{jjspec_bib}

@article{fatemi_synthesizing_2018,
	title = {Synthesizing {Coulombic} superconductivity in van der {Waals} bilayers},
	volume = {98},
	number = {9},
	urldate = {2026-02-12},
	journal = {Physical Review B},
	author = {Fatemi, Valla and Ruhman, Jonathan},
	month = sep,
	year = {2018},
	pages = {094517},
}

@article{hwang09,
  title = {Plasmon modes of spatially separated double-layer graphene},
  author = {Hwang, E. H. and Das Sarma, S.},
  journal = {Phys. Rev. B},
  volume = {80},
  issue = {20},
  pages = {205405},
  numpages = {5},
  year = {2009},
  month = {Nov},
  doi = {10.1103/PhysRevB.80.205405},
  url = {https://link.aps.org/doi/10.1103/PhysRevB.80.205405}
}

@article{bretheau_exciting_2013,
	title = {Exciting {Andreev} pairs in a superconducting atomic contact},
	volume = {499},
	issn = {1476-4687},
	number = {7458},
	urldate = {2022-06-06},
	journal = {Nature},
	author = {Bretheau, L. and Girit, {\c C}. {\" O}. and Pothier, H. and Esteve, D. and Urbina, C.},
	month = jul,
	year = {2013},
    pages = {312--315}
}

@article{khitrova_vacuum_2006,
	title = {Vacuum {Rabi} splitting in semiconductors},
	volume = {2},
	number = {2},
	urldate = {2025-01-31},
	journal = {Nature Physics},
	author = {Khitrova, G. and Gibbs, H. M. and Kira, M. and Koch, S. W. and Scherer, A.},
	month = feb,
	year = {2006},
	pages = {81--90},
}

@article{mermin_lindhard_1970,
	title = {Lindhard {Dielectric} {Function} in the {Relaxation}-{Time} {Approximation}},
	volume = {1},
	number = {5},
	urldate = {2026-01-05},
	journal = {Physical Review B},
	author = {Mermin, N. D.},
	month = mar,
	year = {1970},
	pages = {2362--2363},
}

@article{sun_collective_2020,
	title = {Collective modes and terahertz near-field response of superconductors},
	volume = {2},
	url = {https://link.aps.org/doi/10.1103/PhysRevResearch.2.023413},
	doi = {10.1103/PhysRevResearch.2.023413},
	number = {2},
	urldate = {2024-06-24},
	journal = {Physical Review Research},
	author = {Sun, Zhiyuan and Fogler, M. M. and Basov, D. N. and Millis, Andrew J.},
	month = jun,
	year = {2020},
	pages = {023413},
}

@article{wunsch_dynamical_2006,
	title = {Dynamical polarization of graphene at finite doping},
	volume = {8},
	number = {12},
	urldate = {2023-08-16},
	journal = {New Journal of Physics},
	author = {Wunsch, B. and Stauber, T. and Sols, F. and Guinea, F.},
	month = dec,
	year = {2006},
	pages = {318},
}

@article{sarma_07,
  title = {Many-body interaction effects in doped and undoped graphene: Fermi liquid versus non-Fermi liquid},
  author = {Das Sarma, S. and Hwang, E. H. and Tse, Wang-Kong},
  journal = {Phys. Rev. B},
  volume = {75},
  issue = {12},
  pages = {121406},
  numpages = {4},
  year = {2007},
  month = {Mar},
  doi = {10.1103/PhysRevB.75.121406},
  url = {https://link.aps.org/doi/10.1103/PhysRevB.75.121406}
}

@article{rmp_graphene,
  title = {Electron-Electron Interactions in Graphene: Current Status and Perspectives},
  author = {Kotov, Valeri N. and Uchoa, Bruno and Pereira, Vitor M. and Guinea, F. and Castro Neto, A. H.},
  journal = {Rev. Mod. Phys.},
  volume = {84},
  issue = {3},
  pages = {1067--1125},
  numpages = {0},
  year = {2012},
  month = {Jul},
  doi = {10.1103/RevModPhys.84.1067},
  url = {https://link.aps.org/doi/10.1103/RevModPhys.84.1067}
}

@article{hwang_dielectric_2007,
	title = {Dielectric function, screening, and plasmons in two-dimensional graphene},
	volume = {75},
	number = {20},
	urldate = {2022-06-03},
	journal = {Physical Review B},
	author = {Hwang, E. H. and Das Sarma, S.},
	month = may,
	year = {2007},
	pages = {205418},
}

@article{polini_plasmons_2008,
	title = {Plasmons and the spectral function of graphene},
	volume = {77},
	number = {8},
	urldate = {2026-03-28},
	journal = {Physical Review B},
	author = {Polini, Marco and Asgari, Reza and Borghi, Giovanni and Barlas, Yafis and Pereg-Barnea, T. and MacDonald, A. H.},
	month = feb,
	year = {2008},
	pages = {081411},
}

@article{lundeberg_tuning_2017,
	title = {Tuning quantum nonlocal effects in graphene plasmonics},
	volume = {357},
	number = {6347},
	urldate = {2025-09-28},
	journal = {Science},
	author = {Lundeberg, Mark B. and Gao, Yuanda and Asgari, Reza and Tan, Cheng and Van Duppen, Ben and Autore, Marta and Alonso-González, Pablo and Woessner, Achim and Watanabe, Kenji and Taniguchi, Takashi and Hillenbrand, Rainer and Hone, James and Polini, Marco and Koppens, Frank H. L.},
	month = jul,
	year = {2017},
	pages = {187--191},
}

@article{ni_fundamental_2018,
	title = {Fundamental limits to graphene plasmonics},
	volume = {557},
	number = {7706},
	urldate = {2023-04-18},
	journal = {Nature},
	author = {Ni, G. X. and McLeod, A. S. and Sun, Z. and Wang, L. and Xiong, L. and Post, K. W. and Sunku, S. S. and Jiang, B.-Y. and Hone, J. and Dean, C. R. and Fogler, M. M. and Basov, D. N.},
	month = may,
	year = {2018},
	pages = {530--533},
}

@article{kipp_cavity_2025,
	title = {Cavity electrodynamics of van der {Waals} heterostructures},
	volume = {21},
	number = {12},
	urldate = {2026-03-28},
	journal = {Nature Physics},
	author = {Kipp, Gunda and Bretscher, Hope M. and Schulte, Benedikt and Herrmann, Dorothee and Kusyak, Kateryna and Day, Matthew W. and Kesavan, Sivasruthi and Matsuyama, Toru and Li, Xinyu and Langner, Sara Maria and Hagelstein, Jesse and Sturm, Felix and Potts, Alexander M. and Eckhardt, Christian J. and Huang, Yunfei and Watanabe, Kenji and Taniguchi, Takashi and Rubio, Angel and Kennes, Dante M. and Sentef, Michael A. and Baudin, Emmanuel and Meier, Guido and Michael, Marios H. and McIver, James W.},
	month = dec,
	year = {2025},
	pages = {1926--1933},
}

@article{hofheinz_bright_2011,
	title = {Bright {Side} of the {Coulomb} {Blockade}},
	volume = {106},
	number = {21},
	urldate = {2024-04-07},
	journal = {Physical Review Letters},
	author = {Hofheinz, M. and Portier, F. and Baudouin, Q. and Joyez, P. and Vion, D. and Bertet, P. and Roche, P. and Esteve, D.},
	month = may,
	year = {2011},
	pages = {217005},
}

@article{holst_effect_1994,
	title = {Effect of a {Transmission} {Line} {Resonator} on a {Small} {Capacitance} {Tunnel} {Junction}},
	volume = {73},
	number = {25},
	urldate = {2022-09-23},
	journal = {Physical Review Letters},
	author = {Holst, T. and Esteve, D. and Urbina, C. and Devoret, M. H.},
	month = dec,
	year = {1994},
	pages = {3455--3458},
}

@article{devoret_effect_1990,
	title = {Effect of the electromagnetic environment on the {Coulomb} blockade in ultrasmall tunnel junctions},
	volume = {64},
	number = {15},
	journal = {Physical Review Letters},
	author = {Devoret, M. H. and Esteve, D. and Grabert, H. and Ingold, G.-L. and Pothier, H. and Urbina, C.},
	month = apr,
	year = {1990},
	pages = {1824--1827},
}

@article{ingold_finite-temperature_1991,
	title = {Finite-{Temperature} {Current}-{Voltage} {Characteristics} of {Ultrasmall} {Tunnel} {Junctions}},
	volume = {14},
	number = {4},
	urldate = {2024-06-17},
	journal = {Europhysics Letters},
	author = {Ingold, G.-L. and Grabert, H.},
	month = feb,
	year = {1991},
	pages = {371},
}

@article{griesmar_superconducting_2021,
	title = {Superconducting on-chip spectrometer for mesoscopic quantum systems},
	volume = {3},
	number = {4},
	journal = {Physical Review Research},
	author = {Griesmar, J. and Rodriguez, R. H. and Benzoni, V. and Pillet, J.-D. and Smirr, J.-L. and Lafont, F. and Girit, {\c C}. {\" O}.},
	month = oct,
	year = {2021},
	pages = {043078},
}

@article{torre_acoustic_2019,
	title = {Acoustic plasmons at the crossover between the collisionless and hydrodynamic regimes in two-dimensional electron liquids},
	volume = {99},
	number = {14},
	urldate = {2026-01-14},
	journal = {Physical Review B},
	author = {Torre, Iacopo and Vieira de Castro, Luan and Van Duppen, Ben and Barcons Ruiz, David and Peeters, FranÃ§ois M. and Koppens, Frank H. L. and Polini, Marco},
	month = apr,
	year = {2019},
	pages = {144307},
}

@article{barcons_ruiz_experimental_2023,
	title = {Experimental signatures of the transition from acoustic plasmon to electronic sound in graphene},
	volume = {9},
	number = {39},
	urldate = {2026-03-28},
	journal = {Science Advances},
	author = {Barcons Ruiz, David and Hesp, Niels C.H. and Herzig Sheinfux, Hanan and Ramos Marim{\'o}n, Carlos and Maissen, Curdin Martin and Principi, Alessandro and Asgari, Reza and Taniguchi, Takashi and Watanabe, Kenji and Polini, Marco and Hillenbrand, Rainer and Torre, Iacopo and Koppens, Frank H.L.},
	month = sep,
	year = {2023},
	pages = {eadi0415},
}

@article{potts_-chip_2023,
	title = {On-{Chip} {Time}-{Domain} {Terahertz} {Spectroscopy} of {Superconducting} {Films} below the {Diffraction} {Limit}},
	volume = {23},
	issn = {1530-6984},
	number = {9},
	urldate = {2023-08-01},
	journal = {Nano Letters},
	author = {Potts, Alex M. and Nayak, Abhay K. and Nagel, Michael and Kaj, Kelson and Stamenic, Biljana and John, Demis D. and Averitt, Richard D. and Young, Andrea F.},
	month = may,
	year = {2023},
	pages = {3835--3841},
}

@misc{potts_finite-momentum_2026,
	title = {Finite-momentum {Cooper} plasmons in superconducting terahertz microcavities},
	urldate = {2026-03-28},
	author = {Potts, Alex M. and Michael, Marios H. and Kipp, Gunda and Langner, Sara M. and Bretscher, Hope M. and Stensberg, Jonathan and Kaj, Kelson and Matsuyama, Toru and Day, Matthew W. and Sturm, Felix and Nayak, Abhay K. and Cohen, Liam A. and Zhu, Xiaoyang and Young, Andrea and McIver, James},
	month = jan,
	year = {2026},
	note = {arXiv:2601.10692 [cond-mat]},
}

@article{grimm_bright_2019,
	title = {Bright {On}-{Demand} {Source} of {Antibunched} {Microwave} {Photons} {Based} on {Inelastic} {Cooper} {Pair} {Tunneling}},
	volume = {9},
	number = {2},
	urldate = {2026-03-28},
	journal = {Physical Review X},
	author = {Grimm, A. and Blanchet, F. and Albert, R. and Leppäkangas, J. and Jebari, S. and Hazra, D. and Gustavo, F. and Thomassin, J.-L. and Dupont-Ferrier, E. and Portier, F. and Hofheinz, M.},
	month = apr,
	year = {2019},
	pages = {021016},
}

@article{pines_1,
  title = {A Collective Description of Electron Interactions. {I}. {M}agnetic Interactions},
  author = {Bohm, David and Pines, David},
  journal = {Phys. Rev.},
  volume = {82},
  issue = {5},
  pages = {625--634},
  numpages = {0},
  year = {1951},
  month = {Jun},
  doi = {10.1103/PhysRev.82.625},
  url = {https://link.aps.org/doi/10.1103/PhysRev.82.625}
}

@article{pines_2,
  title = {A Collective Description of Electron Interactions: {II}. {C}ollective $\mathrm{vs}$ Individual Particle Aspects of the Interactions},
  author = {Pines, David and Bohm, David},
  journal = {Phys. Rev.},
  volume = {85},
  issue = {2},
  pages = {338--353},
  numpages = {0},
  year = {1952},
  month = {Jan},
  doi = {10.1103/PhysRev.85.338},
  url = {https://link.aps.org/doi/10.1103/PhysRev.85.338}
}

@article{pines_3,
  title = {A Collective Description of Electron Interactions: {III}. {C}oulomb Interactions in a Degenerate Electron Gas},
  author = {Bohm, David and Pines, David},
  journal = {Phys. Rev.},
  volume = {92},
  issue = {3},
  pages = {609--625},
  numpages = {0},
  year = {1953},
  month = {Nov},
  doi = {10.1103/PhysRev.92.609},
  url = {https://link.aps.org/doi/10.1103/PhysRev.92.609}
}

@book{Mahan1990ManyParticle,
  author    = {G. D. Mahan},
  title     = {Many-Particle Physics},
  edition   = {2},
  year      = {1990},
  publisher = {Plenum Press},
  address   = {New York}
}

@book{pines1989theory,
  title={Theory of Quantum Liquids: Normal Fermi Liquids},
  author={Pines, David and Nozi{\`e}res, Philippe},
  isbn={9780429492662},
  year={1989},
  publisher={CRC Press}
}

@book{Giuliani:826125,
      author        = "Giuliani, Gabriele F and Vignale, Giovanni",
      title         = "{Quantum theory of the electron liquid}",
      publisher     = "Cambridge Univ. Press",
      address       = "Cambridge",
      year          = "2005",
      isbn          = "978-0521826129"
}

@misc{michael_resolving_2025,
	title = {Resolving self-cavity effects in two-dimensional quantum materials},
	urldate = {2025-09-29},
	author = {Michael, Marios H. and Kipp, Gunda and Potts, Alexander M. and Day, Matthew W. and Matsuyama, Toru and Meier, Guido and Bretscher, Hope M. and McIver, James W.},
	month = may,
	year = {2025},
	note = {arXiv:2505.12799 [cond-mat]},
}

@article{schlawin_cavity_2022,
	title = {Cavity quantum materials},
	volume = {9},
	issn = {1931-9401},
	number = {1},
	urldate = {2024-12-21},
	journal = {Applied Physics Reviews},
	author = {Schlawin, F. and Kennes, D. M. and Sentef, M. A.},
	month = feb,
	year = {2022},
	pages = {011312},
}

@article{sentef_cavity_2018,
	title = {Cavity quantum-electrodynamical polaritonically enhanced electron-phonon coupling and its influence on superconductivity},
	volume = {4},
	number = {11},
	urldate = {2026-03-30},
	journal = {Science Advances},
	author = {Sentef, M. A. and Ruggenthaler, M. and Rubio, A.},
	month = nov,
	year = {2018},
	pages = {eaau6969},
}

@article{garcia-vidal_manipulating_2021,
	title = {Manipulating matter by strong coupling to vacuum fields},
	volume = {373},
	number = {6551},
	urldate = {2026-03-30},
	journal = {Science},
	author = {Garcia-Vidal, Francisco J. and Ciuti, Cristiano and Ebbesen, Thomas W.},
	month = jul,
	year = {2021},
	pages = {eabd0336},
}

@article{keren_cavity-altered_2026,
	title = {Cavity-altered superconductivity},
	volume = {650},
	copyright = {2026 The Author(s)},
	issn = {1476-4687},
	number = {8103},
	urldate = {2026-03-30},
	journal = {Nature},
	author = {Keren, Itai and Webb, Tatiana A. and Zhang, Shuai and Xu, Jikai and Sun, Dihao and Kim, Brian S. Y. and Shin, Dongbin and Zhang, Songtian S. and Zhang, Junhe and Pereira, Giancarlo and Yao, Juntao and Okugawa, Takuya and Michael, Marios H. and Viñas Boström, Emil and Edgar, James H. and Wolf, Stuart and Julian, Matthew and Prasankumar, Rohit P. and Miyagawa, Kazuya and Kanoda, Kazushi and Gu, Genda and Cothrine, Matthew and Mandrus, David and Buzzi, Michele and Cavalleri, Andrea and Dean, Cory R. and Kennes, Dante M. and Millis, Andrew J. and Li, Qiang and Sentef, Michael A. and Rubio, Angel and Pasupathy, Abhay N. and Basov, D. N.},
	month = feb,
	year = {2026},
	pages = {864--868},
}

@article{flick_strong_2018,
	title = {Strong light-matter coupling in quantum chemistry and quantum photonics},
	volume = {7},
	copyright = {De Gruyter expressly reserves the right to use all content for commercial text and data mining within the meaning of Section 44b of the German Copyright Act.},
	issn = {2192-8614},
	number = {9},
	urldate = {2026-03-30},
	journal = {Nanophotonics},
	author = {Flick, Johannes and Rivera, Nicholas and Narang, Prineha},
	month = sep,
	year = {2018},
	pages = {1479--1501},
}

@article{basov_polaritons_2016,
	title = {Polaritons in van der {Waals} materials},
	volume = {354},
	number = {6309},
	urldate = {2026-03-30},
	journal = {Science},
	author = {Basov, D. N. and Fogler, M. M. and Garc{\'i}a de Abajo, F. J.},
	month = oct,
	year = {2016},
	pages = {aag1992},
}

@article{low_polaritons_2017,
	title = {Polaritons in layered two-dimensional materials},
	volume = {16},
	copyright = {2016 Springer Nature Limited},
	issn = {1476-4660},
	number = {2},
	urldate = {2026-03-30},
	journal = {Nature Materials},
	author = {Low, Tony and Chaves, Andrey and Caldwell, Joshua D. and Kumar, Anshuman and Fang, Nicholas X. and Avouris, Phaedon and Heinz, Tony F. and Guinea, Francisco and Martin-Moreno, Luis and Koppens, Frank},
	month = feb,
	year = {2017},
	pages = {182--194},
}

@article{appugliese_breakdown_2022,
	title = {Breakdown of topological protection by cavity vacuum fields in the integer quantum {Hall} effect},
	volume = {375},
	number = {6584},
	urldate = {2026-03-30},
	journal = {Science},
	author = {Appugliese, Felice and Enkner, Josefine and Paravicini-Bagliani, Gian Lorenzo and Beck, Mattias and Reichl, Christian and Wegscheider, Werner and Scalari, Giacomo and Ciuti, Cristiano and Faist, JÃ©rÃ´me},
	month = mar,
	year = {2022},
	pages = {1030--1034},
}

@article{jarc_cavity-mediated_2023,
	title = {Cavity-mediated thermal control of metal-to-insulator transition in {1T}-{TaS$_2$}},
	volume = {622},
	copyright = {2023 The Author(s), under exclusive licence to Springer Nature Limited},
	issn = {1476-4687},
	number = {7983},
	urldate = {2026-03-30},
	journal = {Nature},
	author = {Jarc, Giacomo and Mathengattil, Shahla Yasmin and Montanaro, Angela and Giusti, Francesca and Rigoni, Enrico Maria and Sergo, Rudi and Fassioli, Francesca and Winnerl, Stephan and Dal Zilio, Simone and Mihailovic, Dragan and Prelovšek, Peter and Eckstein, Martin and Fausti, Daniele},
	month = oct,
	year = {2023},
	pages = {487--492},
}

@article{thomas_large_2021,
	title = {Large {Enhancement} of {Ferromagnetism} under a {Collective} {Strong} {Coupling} of {YBCO} {Nanoparticles}},
	volume = {21},
	issn = {1530-6984},
	number = {10},
	urldate = {2026-03-30},
	journal = {Nano Letters},
	author = {Thomas, Anoop and Devaux, Eloise and Nagarajan, Kalaivanan and Rogez, Guillaume and Seidel, Marcus and Richard, Fanny and Genet, Cyriaque and Drillon, Marc and Ebbesen, Thomas W.},
	month = may,
	year = {2021},
	pages = {4365--4370},
}

@article{lu_cavity-enhanced_2024,
	title = {Cavity-enhanced superconductivity in {MgB$_2$} from first-principles quantum electrodynamics ({QEDFT})},
	volume = {121},
	number = {50},
	urldate = {2026-03-30},
	journal = {Proceedings of the National Academy of Sciences},
	author = {Lu, I-Te and Shin, Dongbin and Svendsen, Mark Kamper and Hübener, Hannes and De Giovannini, Umberto and Latini, Simone and Ruggenthaler, Michael and Rubio, Angel},
	month = dec,
	year = {2024},
	pages = {e2415061121},
}

@article{hubener_quantum_2024,
	title = {Quantum materials engineering by structured cavity vacuum fluctuations},
	volume = {4},
	issn = {2633-4356},
	number = {2},
	urldate = {2026-03-30},
	journal = {Materials for Quantum Technology},
	author = {H{\"u}bener, Hannes and Bostr{\"o}m, Emil Vi{\~n}as and Claassen, Martin and Latini, Simone and Rubio, Angel},
	month = jun,
	year = {2024},
	pages = {023002},
}

@misc{xu_vacuum-dressed_2026,
	title = {Vacuum-dressed superconductivity in {NbN} observed in a high-{$Q$} terahertz cavity},
	url = {https://arxiv.org/abs/2601.08191v1},
	urldate = {2026-03-30},
	journal = {arXiv.org},
	author = {Xu, Hongjing and Baydin, Andrey and Yi, Qinyan and Lu, I.-Te and Zhu, Ningxu and Kritzell, T. Elijah and Doumani, Jacques and Kim, Dasom and Tay, Fuyang and Rubio, Angel and Kono, Junichiro},
	month = jan,
	year = {2026},
    note = {arXiv:2603.08191 [cond-mat]}
}

@article{li_vacuum_2018,
	title = {Vacuum {Bloch}–{Siegert} shift in {Landau} polaritons with ultra-high cooperativity},
	volume = {12},
	copyright = {2018 The Author(s)},
	issn = {1749-4893},
	number = {6},
	urldate = {2024-01-04},
	journal = {Nature Photonics},
	author = {Li, Xinwei and Bamba, Motoaki and Zhang, Qi and Fallahi, Saeed and Gardner, Geoff C. and Gao, Weilu and Lou, Minhan and Yoshioka, Katsumasa and Manfra, Michael J. and Kono, Junichiro},
	month = jun,
	year = {2018},
	pages = {324--329},
}

@article{eckhardt_surface-mediated_2025,
	title = {Surface-{Mediated} {Ultrastrong} {Cavity} {Coupling} of {Two}-{Dimensional} {Itinerant} {Electrons}},
	volume = {135},
	number = {15},
	urldate = {2026-03-31},
	journal = {Physical Review Letters},
	author = {Eckhardt, Christian J. and Grankin, Andrey and Kennes, Dante M. and Ruggenthaler, Michael and Rubio, Angel and Sentef, Michael A. and Hafezi, Mohammad and Michael, Marios H.},
	month = oct,
	year = {2025},
	pages = {156902},
}

@article{gallagher_quantum-critical_2019,
	title = {Quantum-critical conductivity of the {Dirac} fluid in graphene},
	volume = {364},
	issn = {0036-8075, 1095-9203},
	number = {6436},
	urldate = {2021-08-14},
	journal = {Science},
	author = {Gallagher, Patrick and Yang, Chan-Shan and Lyu, Tairu and Tian, Fanglin and Kou, Rai and Zhang, Hai and Watanabe, Kenji and Taniguchi, Takashi and Wang, Feng},
	month = apr,
	year = {2019},
	pmid = {30819930},
	pages = {158--162},
}

@article{sun_universal_2018,
	title = {Universal linear and nonlinear electrodynamics of a {Dirac} fluid},
	volume = {115},
	issn = {0027-8424, 1091-6490},
	number = {13},
	urldate = {2022-01-05},
	journal = {Proceedings of the National Academy of Sciences},
	author = {Sun, Zhiyuan and Basov, Dmitry N. and Fogler, Michael M.},
	month = mar,
	year = {2018},
	pages = {3285--3289},
}

@misc{bluefors_note,
    title={The cooling powers at 20 {mK} are the range of quoted {B}luefors {LD}- and {XLD}-series dilution refrigerator performance}
}

@article{passetti_cavity_2023,
	title = {Cavity {Light}-{Matter} {Entanglement} through {Quantum} {Fluctuations}},
	volume = {131},
	url = {https://link.aps.org/doi/10.1103/PhysRevLett.131.023601},
	doi = {10.1103/PhysRevLett.131.023601},
	number = {2},
	urldate = {2026-04-06},
	journal = {Physical Review Letters},
	author = {Passetti, Giacomo and Eckhardt, Christian J. and Sentef, Michael A. and Kennes, Dante M.},
	month = jul,
	year = {2023},
	pages = {023601},
}

@article{alonso-gonzalez_acoustic_2017,
	title = {Acoustic terahertz graphene plasmons revealed by photocurrent nanoscopy},
	volume = {12},
	copyright = {2016 Springer Nature Limited},
	issn = {1748-3395},
	url = {https://www.nature.com/articles/nnano.2016.185},
	doi = {10.1038/nnano.2016.185},
	number = {1},
	urldate = {2026-04-07},
	journal = {Nature Nanotechnology},
	author = {Alonso-Gonz{\'a}lez, Pablo and Nikitin, Alexey Y. and Gao, Yuanda and Woessner, Achim and Lundeberg, Mark B. and Principi, Alessandro and Forcellini, NicolÃ² and Yan, Wenjing and V{\'e}lez, Sa{\"u}l and Huber, Andreas J. and Watanabe, Kenji and Taniguchi, Takashi and Casanova, FÃ©lix and Hueso, Luis E. and Polini, Marco and Hone, James and Koppens, Frank H. L. and Hillenbrand, Rainer},
	month = jan,
	year = {2017},
	pages = {31--35},
}

@article{barlas_chirality_2007,
	title = {Chirality and {Correlations} in {Graphene}},
	volume = {98},
	url = {https://link.aps.org/doi/10.1103/PhysRevLett.98.236601},
	doi = {10.1103/PhysRevLett.98.236601},
	number = {23},
	urldate = {2025-12-29},
	journal = {Physical Review Letters},
	author = {Barlas, Yafis and Pereg-Barnea, T. and Polini, Marco and Asgari, Reza and MacDonald, A. H.},
	month = jun,
	year = {2007},
	pages = {236601},
}

@article{das_sarma_velocity_2013,
	title = {Velocity renormalization and anomalous quasiparticle dispersion in extrinsic graphene},
	volume = {87},
	url = {https://link.aps.org/doi/10.1103/PhysRevB.87.045425},
	doi = {10.1103/PhysRevB.87.045425},
	number = {4},
	urldate = {2026-04-07},
	journal = {Physical Review B},
	author = {Das Sarma, S. and Hwang, E. H.},
	month = jan,
	year = {2013},
	pages = {045425},
}

@article{abedinpour_drude_2011,
	title = {Drude weight, plasmon dispersion, and ac conductivity in doped graphene sheets},
	volume = {84},
	url = {https://link.aps.org/doi/10.1103/PhysRevB.84.045429},
	doi = {10.1103/PhysRevB.84.045429},
	number = {4},
	urldate = {2025-12-29},
	journal = {Physical Review B},
	author = {Abedinpour, Saeed H. and Vignale, G. and Principi, A. and Polini, Marco and Tse, Wang-Kong and MacDonald, A. H.},
	month = jul,
	year = {2011},
	pages = {045429},
}

@article{ciuti_quantum_2005,
	title = {Quantum vacuum properties of the intersubband cavity polariton field},
	volume = {72},
	url = {https://link.aps.org/doi/10.1103/PhysRevB.72.115303},
	doi = {10.1103/PhysRevB.72.115303},
	number = {11},
	urldate = {2026-04-17},
	journal = {Physical Review B},
	author = {Ciuti, Cristiano and Bastard, Gérald and Carusotto, Iacopo},
	month = sep,
	year = {2005},
	pages = {115303},
}

@article{kennes_moire_2021,
	title = {Moir{\'e} heterostructures as a condensed-matter quantum simulator},
	volume = {17},
	url = {https://www.nature.com/articles/s41567-020-01154-3},
	doi = {10.1038/s41567-020-01154-3},
	number = {2},
	urldate = {2022-06-02},
	journal = {Nature Physics},
	author = {Kennes, Dante M. and Claassen, Martin and Xian, Lede and Georges, Antoine and Millis, Andrew J. and Hone, James and Dean, Cory R. and Basov, D. N. and Pasupathy, Abhay N. and Rubio, Angel},
	month = feb,
	year = {2021},
	pages = {155--163},
}

@article{mak_semiconductor_2022,
	title = {Semiconductor moiré materials},
	volume = {17},
	issn = {1748-3395},
	url = {https://www.nature.com/articles/s41565-022-01165-6},
	doi = {10.1038/s41565-022-01165-6},
	number = {7},
	urldate = {2023-08-21},
	journal = {Nature Nanotechnology},
	author = {Mak, Kin Fai and Shan, Jie},
	month = jul,
	year = {2022},
	pages = {686--695},
}

@article{enkner_tunable_2025,
	title = {Tunable vacuum-field control of fractional and integer quantum {Hall} phases},
	volume = {641},
	url = {https://www.nature.com/articles/s41586-025-08894-3},
	doi = {10.1038/s41586-025-08894-3},
	number = {8064},
	urldate = {2026-03-30},
	journal = {Nature},
	author = {Enkner, Josefine and Graziotto, Lorenzo and Bori{\c{c}}i, Dalin and Appugliese, Felice and Reichl, Christian and Scalari, Giacomo and Regnault, Nicolas and Wegscheider, Werner and Ciuti, Cristiano and Faist, J{\'e}r{\^o}me},
	month = may,
	year = {2025},
	pages = {884--889},
}

@misc{helmrich_cavity-driven_2025,
	title = {Cavity-{Driven} {Attractive} {Interactions} in {Quantum} {Materials}},
	url = {http://arxiv.org/abs/2408.00189},
	doi = {10.48550/arXiv.2408.00189},
	urldate = {2026-04-22},
	author = {Helmrich, F. and Adlong, H. S. and Khanonkin, I. and Kroner, M. and Scalari, G. and Faist, J. and Imamoglu, A. and Nova, T. F.},
	month = apr,
	year = {2025},
	note = {arXiv:2408.00189 [cond-mat]},
}

@misc{chen_terahertz_2025,
	title = {Terahertz electrodynamics in a zero-field {Wigner} crystal},
	url = {http://arxiv.org/abs/2509.10624},
	doi = {10.48550/arXiv.2509.10624},
	urldate = {2025-10-24},
	publisher = {arXiv},
	author = {Chen, Su-Di and Qi, Ruishi and Kim, Ha-Leem and Feng, Qixin and Xia, Ruichen and Abeysinghe, Dishan and Xie, Jingxu and Taniguchi, Takashi and Watanabe, Kenji and Lee, Dung-Hai and Wang, Feng},
	month = sep,
	year = {2025},
}

@article{seo_-chip_2024,
	title = {On-{Chip} {Terahertz} {Spectroscopy} for {Dual}-{Gated} van der {Waals} {Heterostructures} at {Cryogenic} {Temperatures}},
	volume = {24},
	issn = {1530-6984},
	url = {https://doi.org/10.1021/acs.nanolett.4c04137},
	doi = {10.1021/acs.nanolett.4c04137},
	number = {47},
	urldate = {2024-12-21},
	journal = {Nano Letters},
	publisher = {American Chemical Society},
	author = {Seo, Junseok and Lu, Zhengguang and Park, Seunghyun and Yang, Jixiang and Xia, Fangzhou and Ye, Shenyong and Yao, Yuxuan and Han, Tonghang and Shi, Lihan and Watanabe, Kenji and Taniguchi, Takashi and Yacoby, Amir and Ju, Long},
	month = nov,
	year = {2024},
	pages = {15060--15067},
}

@article{blagg_-chip_2022,
	title = {On-chip heating effects in electronic measurements at cryogenic temperatures},
	volume = {126},
	issn = {0011-2275},
	url = {https://www.sciencedirect.com/science/article/pii/S0011227522001187},
	doi = {10.1016/j.cryogenics.2022.103536},
	urldate = {2026-05-16},
	journal = {Cryogenics},
	author = {Blagg, Kirsten and Castagn{\`e}de, Antoine and Singh, Meenakshi},
	month = sep,
	year = {2022},
	pages = {103536},
}

@misc{bon-mardion_sub-kelvin_2026,
	title = {Sub-kelvin thermal conductivity of substrates and on-chip routing in quantum integrated systems},
	url = {http://arxiv.org/abs/2605.06146},
	doi = {10.48550/arXiv.2605.06146},
	urldate = {2026-05-16},
	publisher = {arXiv},
	author = {Bon-Mardion, Charles and Lorin, Arnaud and Deschaseaux, Edouard and Feautrier, C{\'e}line and Mermin, Daniel and Charbonnier, Jean and Li, Jing and Sauvageot, Jean-Luc and Thomas, Candice},
	month = may,
	year = {2026},
	note = {arXiv:2605.06146 [cond-mat.mes-hall]},
}

@article{yu_interaction_2013,
	title = {Interaction phenomena in graphene seen through quantum capacitance},
	volume = {110},
	url = {https://www.pnas.org/doi/full/10.1073/pnas.1300599110},
	doi = {10.1073/pnas.1300599110},
	number = {9},
	urldate = {2026-05-16},
	journal = {Proceedings of the National Academy of Sciences},
	publisher = {Proceedings of the National Academy of Sciences},
	author = {Yu, G. L. and Jalil, R. and Belle, Branson and Mayorov, Alexander S. and Blake, Peter and Schedin, Frederick and Morozov, Sergey V. and Ponomarenko, Leonid A. and Chiappini, F. and Wiedmann, S. and Zeitler, Uli and Katsnelson, Mikhail I. and Geim, A. K. and Novoselov, Kostya S. and Elias, Daniel C.},
	month = feb,
	year = {2013},
	pages = {3282--3286},
}

@misc{bretscher_fluctuation_2026,
	title = {Fluctuation engineering in cavity quantum materials},
	url = {http://arxiv.org/abs/2604.08666},
	doi = {10.48550/arXiv.2604.08666},
	urldate = {2026-05-18},
	publisher = {arXiv},
	author = {Bretscher, Hope M. and Graziotto, Lorenzo and Michael, Marios H. and Montanaro, Angela and Lu, I.-Te and Grankin, Andrey and McIver, James W. and Faist, Jerome and Fausti, Daniele and Eckstein, Martin and Ruggenthaler, Michael and Rubio, Angel and Basov, D. N. and Hafezi, Mohammad and Claassen, Martin and Kennes, Dante M. and Sentef, Michael A.},
	month = apr,
	year = {2026},
	note = {arXiv:2604.08666 [cond-mat.mes-hall]},
}

@article{deblock_detection_2003,
	title = {Detection of {Quantum} {Noise} from an {Electrically} {Driven} {Two}-{Level} {System}},
	volume = {301},
	url = {https://www.science.org/doi/10.1126/science.1084175},
	doi = {10.1126/science.1084175},
	number = {5630},
	urldate = {2026-05-19},
	journal = {Science},
	publisher = {American Association for the Advancement of Science},
	author = {Deblock, Richard and Onac, Eugen and Gurevich, Leonid and Kouwenhoven, Leo P.},
	month = jul,
	year = {2003},
	pages = {203--206},
}

@article{peugeot_two-tone_2024,
	title = {Two-tone spectroscopy of high-frequency quantum circuits with a {Josephson} emitter},
	volume = {22},
	url = {https://link.aps.org/doi/10.1103/PhysRevApplied.22.064027},
	doi = {10.1103/PhysRevApplied.22.064027},
	number = {6},
	urldate = {2026-05-20},
	journal = {Physical Review Applied},
	publisher = {American Physical Society},
	author = {Peugeot, A. and Riechert, H. and Annabi, S. and Balembois, L. and Villiers, M. and Flurin, E. and Griesmar, J. and Arrighi, E. and Pillet, J.-D. and Bretheau, L.},
	month = dec,
	year = {2024},
	pages = {064027},
}

\end{document}